\documentstyle[twocolumn,eqsecnum,pra,aps,epsf]{revtex}
\begin{document}
\draft
\title{Casimir force in absorbing multilayers}
\author{M. S. Toma\v s}
\address{Rudjer Bo\v skovi\' c Institute, P. O. B. 1016, 10002 Zagreb,
Croatia}
\date{July 18, 2002}
\maketitle
\begin{abstract}
The Casimir effect in a dispersive and absorbing multilayered system 
is considered adopting the (net) vacuum-field pressure point of view 
to the Casimir force. Using the properties of the macroscopic field 
operators appropriate for absorbing systems and a convenient compact 
form of the Green function for a multilayer, a straightforward and 
transparent derivation of the Casimir force in a lossless layer of an
otherwise absorbing multilayer is presented. The resulting expression 
in terms of the reflection coefficients of the surrounding stacks 
of layers is of the same form as that obtained by Zhou and Spruch 
for a purely dispersive multilayer using the (surface) mode summation 
method  [Phys. Rev. A {\bf 52}, 297 (1995)]. Owing to the recursion 
relations which the generalized Fresnel coefficients satisfy, this 
result can be applied to more complex systems with planar symmetry. 
This is illustrated by calculating the Casimir force on a dielectric 
(metallic) slab in a planar cavity with realistic mirrors. Also, a 
relationship between the Casimir force and energy in two different 
layers is established.
\end{abstract}

\section{Introduction}
Originally, the Casimir effect was predicted as a feature of the 
electromagnetic field between two neutral ideally conducting 
plates and consisted in the appearance of an attractive force 
between the plates. The force is due to the change of the 
zero-point energy of the field in the confined space \cite{casimir}. 
In this special case, however, the Casimir force can also be 
viewed as the long-range van der Waals force
\cite{lifshitz0,schwinger,milonni}. It becomes appreciable 
in the submicron and rapidly increases in the nanometer range. 
As such, it may strongly affect processing in nanotechnology 
as well as functioning of micro- and nanomachines and devices 
\cite{bordag}. Clearly, these new developments pose the problem
of realistic calculations of the Casimir force on objects
in complex enviroments.

In contrast to the highly idealized system considered by 
Casimir \cite{casimir}, Lifshitz \cite{lifshitz0} calculated 
the force between two thick (semi-infinite) dielectric slabs 
by taking into account the dispersion and absorption in 
the dielectrics as well as the temperature effects. In this 
respect, his theory is far more realistic and, as the effects 
of finite conductivity and dissipation in the metal 
can be observed in the recent high-precision experiments 
\cite{lamoreaux,mohideen,bressi}, his result for the force
at zero temperature is standardly used when analyzing the Casimir 
force in the planar geometry \cite{lamoreaux1,lambrecht}. The
Lifshitz approach is based on the calculation of the electromagnetic 
field due to the randomly fluctuating currents in the dielectric 
slabs and on the subsequent calculation of the Maxwell stress 
tensor in the region inbetween. Owing to its complexity, however, 
it has never been extended to the calculation of the force between  
multilayered stacks although the generalization of the final
result to this configuration is fairly obvious.

The Casimir effect in multilayered systems is usually considered 
using either the surface mode summation method
\cite{kampen,schram,zhou,klim} (see also Refs. \cite{milonni,bordag})
to calculate the change in the electromagnetic field zero-point 
energy due to the presence of the dielectric stacks, or the stress 
tensor method \cite{schwinger,brown} to calculate directly
the vacuum field pressure on the stacks. Strictly speaking,  
the mode summation method applies only to purely dispersive 
(lossless) systems as only in this case the mode frequencies 
are real. However, when expressed as an integral over the 
imaginary frequency, the final result for the Casimir energy 
(force) turns out to be applicable to absorbing systems as well.
An indication that this must be so is the fact that the dielectric 
function is always real on the imaginary axis irrespective of 
whether the system is absorbing or not \cite{milonni}. Thus, while 
in their calculation of the force in a multilayer Zhou 
and Spruch \cite{zhou} assumed a purely dispersive system,
Klimchetskaya {\it et al}. \cite{klim} recently considered
a similiar but absorbing system. 

On the other hand, being a local approach, the stress tensor
method does not necessarily imply a lossless system. Since
the stress tensor cannot be defined macroscopically for an 
absorbing medium \cite{ll}, the only necessary assumption is 
actually that the region where the vacuum-field pressure is 
calculated is nonabsorbing, whereas the other parts of the 
system may generally be dissipative. Despite this fact, 
numerous papers in the past used the stress tensor method to 
calculate the Casimir force assuming, at most, a dispersive 
but nonabsorbing system. One of the reasons for that is 
certainly lack of knowledge on the proper form and properties
of macroscopic field operators aproppriate for an absorbing 
system at that time. 

The first calculation of the Casimir force between two absorbing 
slabs is due to Kupiszewska \cite{kupi1} who modeled dielectric 
atoms as a collection of harmonic oscillators coupled to a heat 
bath that absorbs energy. Only the modes propagating normally 
to the slabs were considered, so that this approach was 
effectively one-dimensional (1D). Describing the reservoir 
through a damping constant and the Langevin force and solving 
for the field operators, Kupiszewska obtained for the 
force between the slabs the same expression in terms of their 
reflection coefficients as that obtained previously for an inert 
\cite{kupi2} or a lossless \cite{jaekel} 1D system, except that 
this time the dielectric function of the slabs was complex. 
Recently, this result was rederived using a Green function method 
for quantizing the macroscopic field in (1D) absorbing systems in 
conjuction with a scattering matrix approach \cite{matloob} and 
was also extended to two identical absorbing superlattices 
\cite{esqu1}. Very recently, Esquivel-Sirvent {\it et al}. 
demonstrated an alternative Green function approach that makes 
the quantization of the field within the slabs unnecessary and 
calculated the Casimir force in an asymmetric configuration 
\cite{esqu2} which was earlier considered only in the lossless 
case \cite{jaekel}.

Owing to their complex structure, an explicit calculation of the 
field operators as in Refs. \cite{kupi1,kupi2,jaekel,matloob}
is highly impractical in the general case of a three-dimensional
(3D) dissipative inhomogeneous system. However, as pointed 
recently by Matloob \cite{matloob2} (see also Ref.\cite{matloob}), 
using the fluctuation-dissipation theorem and the linear response 
theory, the field correlation functions needed to calculate the 
stress tensor can be expressed in terms of the (classical) Green 
function for the system. 
In this way, only the knowledge of the Green function is, therefore, 
actually needed to calculate the Casimir force. Using this method, 
Matloob and Falinejad recently considered the Casimir force between 
two identical absorbing dielectric slabs \cite{matloob3}. Very 
recently Moch\'{a}n {\it et al}. 
\cite{mochan} generalized their Green function method \cite{esqu2} 
to three dimensions and calculated the Casimir force between two 
arbitrary slabs. Expressing the reflection coefficients of the 
slabs through the generalized surface impedances, these authors 
argued that their formal result could be applied to rather 
general but not chiral media, also including nonlocal 
inhomogeneous dissipative slabs. In Refs. \cite{matloob3,mochan} 
the space between the slabs was assumed empty. 

In this work we calculate the Casimir force in a lossless  
dispersive layer of an otherwise absorbing multilayer by employing 
the macroscopic field operators as emerge from a recently developed 
scheme for quantizing the electromagnetic field in inhomogeneous 
dissipative 3D-systems \cite{welsch,matloob1} and using a convenient 
Green function for a multilayer \cite{tomas}. In this way, we obtain 
a general result for the Casimir force in stratified local media. 
In addition, using the properties of the 
generalized Fresnel coefficients, we derive a relationship between 
the Casimir force and energy in two different layers and demonstrate 
the applicability of the theory to more complex planar systems by 
calculating the Casimir force on a diectric slab in a realistic 
planar cavity.

\section{Theory}
Consider a multilayered system described by the dielectric 
function $\varepsilon({\bf r},\omega)=\varepsilon'({\bf r},\omega)+
i\varepsilon''({\bf r},\omega)$ defined in a  stepwise 
fashion, as depicted in Fig. 1. The Casimir force in a layer 
corresponds to the net vacuum-field pressure in the multilayer 
with respect to the pressure in the infinite layer (medium). 
Accordingly, the force ${\bf F}_{j/l}$ on a stack of layers 
that separates a $j$th and an $l$th layer is given by
\cite{jackson}
\begin{equation}
{\bf F}_{j/l}=Af_{j/l}\hat{\bf z},\;\;\;
f_{j/l}=\pm(\tilde{T}_{l,zz}-\tilde{T}_{j,zz}),
\label{flj}
\end{equation}
where $\tilde{T}_{j,zz}$ is the $zz$ component of the regularized 
stress tensor in the $j$th layer
\begin{equation}
\tilde{T}_{j,ab}=T_{j,ab}-T^0_{j,ab},
\end{equation}
with $T_{j,ab}$ and $T^0_{j,ab}$ being the corresponding Maxwell 
stress tensors in the multilayer and in the infinite 
medium ($j$), respectively. In Eq. (\ref{flj}), $A$ is the 
area of the stack and the $+$ ($-$) sign applies if $l>j$ ($l<j$).
Since the regularized stress tensor vanishes in the outmost layers, 
we have for $f_{j-}\equiv f_{j/0}$ and $f_{j+}\equiv f_{j/n}$: 
\begin{equation}
f_{j-}=-f_{j+}=\tilde{T}_{j,zz},
\label{f}
\end{equation}
so that $\tilde{T}_{j,zz}$ coincides with the force per unit area 
acting on the left (right) stack of layers bounding the layer ($j$). 
\begin{figure}[htb]
\begin{center}
\leavevmode
\hbox{%
\epsfxsize=8.6cm
\epsffile{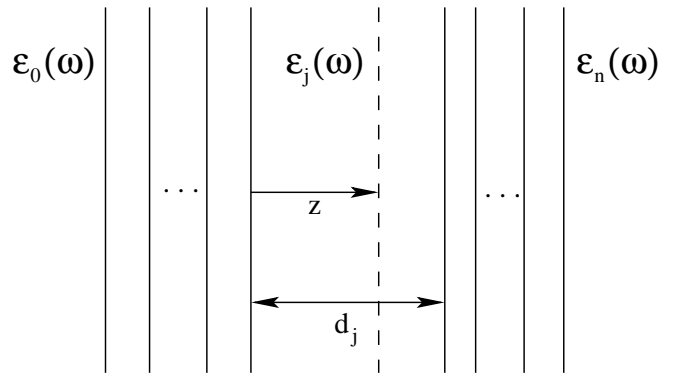}}
\end{center}
\caption{System considered schematically. The dashed line 
represents the plane where the stress tensor is calculated.}
\end{figure}

Replacing field variables in the classical Maxwell stress 
tensor \cite{ll} by the corresponding Heisenberg operators and 
taking its average \cite{brevik}, $T_{j,zz}$ in a lossless layer 
$(j)$ [$\varepsilon''_j(\omega)=0$] is given by
\begin{equation}
T_{j,zz}=\frac{1}{8\pi}\left<E_zD_z-
{\bf E}_\parallel\cdot{\bf D}_\parallel+B_zH_z-
{\bf B}_\parallel\cdot{\bf H}_\parallel\right>_{{\bf r}\in(j)},
\label{T}
\end{equation}
where we have suppressed the argument $({\bf r},t)$ of the field 
operators and the brackets denote the expectation value in the 
vacuum state of the field. In order to calculate the correlation 
functions that appear in Eq. (\ref{T}), we use the properties of 
the macroscopic field operators appropriate for absorbing 
systems \cite{welsch}. These operators are decomposed into their 
"annihilation" and "creation" components according to
\begin{equation}
{\bf E}({\bf r},t)=\int_0^\infty d\omega
{\bf E}({\bf r},\omega)e^{-i\omega t}+H. c.
\label{E}
\end{equation}
and, with the constitutive relations
\[{\bf D}({\bf r},\omega)=
\varepsilon({\bf r},\omega){\bf E}({\bf r},\omega)+
4\pi{\bf P}_N({\bf r},\omega),\]
\begin{equation}
{\bf B}({\bf r},\omega)={\bf H}({\bf r},\omega),
\label{CE}
\end{equation}
obey the standard macroscopic Maxwell equations. Here 
${\bf P}_N({\bf r},\omega)$ and ${\bf P}^\dagger_N({\bf r},\omega)$ 
are the noise polarization operators related to the dissipation 
in the system and obeying the commutation rules
(in the dyadic form): 
\begin{equation}
[{\bf P}_N({\bf r},\omega),{\bf P}^\dagger_N({\bf r}',\omega')]
=\frac{\hbar\varepsilon''({\bf r},\omega)}{4\pi^2}
\tensor{\bf I}\delta({\bf r}-{\bf r}')\delta(\omega-\omega'),
\label{PP}
\end{equation}
where $\tensor{\bf I}$ is the unit dyadic. Therefore, any 
(annihilation) field operator is related 
to ${\bf P}_N({\bf r},\omega)$ via the classical Green function 
$\tensor{\bf G}({\bf r},{\bf r'};\omega)$ satisfying
\begin{equation}
\left[\nabla\times\nabla\times-\varepsilon({\bf r}.\omega)
\frac{\omega^2}{c^2}\tensor{\bf I}\cdot\right]
\tensor{\bf G}({\bf r},{\bf r'};\omega)=4\pi\tensor{\bf I}
\delta({\bf r}-{\bf r'})
\label{GF}
\end{equation}
according to
\begin{equation}
{\bf E}({\bf r},\omega)=\frac{\omega^2}{c^2}\int d^3{\bf r}'
\tensor{\bf G}({\bf r},{\bf r'};\omega)\cdot{\bf P}_N({\bf r}',\omega).
\end{equation}
As a consequence, all field correlation functions can be 
expressed through the Green function in accordance with the 
fluctuation-dissipation theorem \cite{lifshitz}. In particular, 
for the electric-field correlation function we have \cite{welsch} 
\begin{equation}
\left<{\bf E}({\bf r},\omega){\bf E}^\dagger({\bf r}',\omega')\right>
=\frac{\hbar}{\pi}\frac{\omega^2}{c^2}
{\rm Im}\tensor{\bf G}({\bf r},{\bf r'};\omega)\delta(\omega-\omega'),
\label{EE}
\end{equation}
and the magnetic-field correlation function is easily obtained 
from this expression using ${\bf B}({\bf r},\omega)=
(-ic/\omega)\nabla\times{\bf E}({\bf r},\omega)$.

Applying the above results to the $j$th layer and taking
into account that $\varepsilon_j(\omega)$ is real and
that ${\bf P}_N({\bf r},\omega)=0$ in this region, we find 
for the relevant correlation functions in Eq. (\ref{T}):
\begin{mathletters}
\label{CF}
\begin{equation}
\left<{\bf E}({\bf r},t){\bf D}({\bf r},t)\right>_{{\bf r}\in(j)}
=\frac{\hbar}{\pi}\int_0^\infty d\omega \tilde{k}^2_j(\omega)
{\rm Im}\tensor{\bf G}_j({\bf r},{\bf r};\omega),
\label{ED}
\end{equation}
\begin{equation}
\left<{\bf B}({\bf r},t){\bf H}({\bf r},t)\right>_{{\bf r}\in(j)}
=\frac{\hbar}{\pi}\int_0^\infty d\omega
{\rm Im}\tensor{\bf G}^B_j({\bf r},{\bf r};\omega),
\label{BH}
\end{equation}
\end{mathletters}
where $\tilde{k}_j(\omega)=\sqrt{\varepsilon_j(\omega)}\omega/c$ 
is the wave vector in the layer,
$\tensor{\bf G}_j({\bf r},{\bf r'};\omega)$ is the Green function 
element for ${\bf r}$ and ${\bf r'}$ both in the layer $(j)$, and
\begin{equation} 
\tensor{\bf G}^B_j({\bf r},{\bf r'};\omega)=
\nabla\times\tensor{\bf G}_j({\bf r},{\bf r'};\omega)\times
\stackrel{\leftarrow}{\nabla'}
\label{GB}
\end{equation}
is the corresponding Green function element for the magnetic field . 
With the above equations inserted in Eq. \ (\ref{T}), the stress 
tensor $T_{j,zz}$ is expressed entirely in terms of the Green function 
and its derivatives analogously to Eq. (\ref{Tr}) below. Similarly,
through Eqs. (\ref{T}) and (\ref{CF}) applied to the infinite medium 
($j$), the stress tensor $T^0_{j,zz}$ is given by the same expression 
with the infinite-medium Green function
$\tensor{\bf G}^0_j({\bf r},{\bf r'};\omega)$. Therefore, 
the regularized stress tensor $\tilde{T}_{j,zz}$ is expressed as
\begin{eqnarray}
\tilde{T}_{j,zz}&=&\frac{\hbar}{4\pi}{\rm Im}\int_0^\infty
\frac{d\omega}{2\pi}\left\{\tilde{k}^2_j(\omega)\times\right.\nonumber\\
&&\left[G^{\rm sc}_{j,zz}({\bf r},{\bf r};\omega)-
G^{\rm sc}_{j,\parallel}({\bf r},{\bf r};\omega)\right]
\nonumber\\
&&\left.+G^{B,\rm sc}_{j,zz}({\bf r},{\bf r};\omega)-
G^{B,\rm sc}_{j,\parallel}({\bf r},{\bf r};\omega)\right\},
\label{Tr}
\end{eqnarray}
where
\begin{equation}
\tensor{\bf G}^{\rm sc}_j({\bf r},{\bf r'};\omega)=
\tensor{\bf G}_j({\bf r},{\bf r'};\omega)-
\tensor{\bf G}^0_j({\bf r},{\bf r'};\omega)
\label{Gsc}
\end{equation}
is the Green function for the scattered field in the $j$th layer
and $G^{\rm sc}_{j,\parallel}({\bf r},{\bf r}';\omega)=
G^{\rm sc}_{j,xx}({\bf r},{\bf r}';\omega)+
G^{\rm sc}_{j,yy}({\bf r},{\bf r}';\omega)$ is its
parallel trace. 

A convenient form of 
$\tensor{\bf G}^{\rm sc}_j({\bf r},{\bf r'};\omega)$ 
for a general multilayer is derived in Ref. \cite{tomas}. In the 
Appendix we quote this Green function and calculate the expression 
in the curly brackets of Eq. (\ref{Tr}). Inserting Eq. (\ref{cb}), 
we find
\begin{eqnarray} 
\tilde{T}_{j,zz}&=&-\frac{\hbar}{\pi}{\rm Re}\int_0^\infty d\omega
\int\frac{d^2{\bf k}}{(2\pi)^2}\beta_j
\sum_{q=p,s}\frac{1-D_{qj}(\omega,k)}{D_{qj}(\omega,k)}\nonumber\\
&=&\frac{\hbar}{2\pi^2}\int_0^\infty d\xi
\int^\infty_0 dkk\kappa_j\sum_{q=p,s}
\frac{1-D_{qj}(i\xi,k)}{D_{qj}(i\xi,k)},
\label{Trf}
\end{eqnarray}
where $\beta_j(\omega,k)=\sqrt{\tilde{k}_j^2(\omega)-k^2}$,
\begin{equation}
D_{qj}(\omega,k)=1-r^q_{j-}(\omega,k) r^q_{j+}(\omega,k) 
e^{2i\beta_j d_j},
\label{Dj}
\end{equation}
and $r^q_{j\pm}(\omega,k)$ are the reflection coefficients of 
the right and left stack of layers bounding the $j$th layer. The 
second line in Eq. (\ref{Trf}) has been obtained by converting the 
integral over the real $\omega$-axis to one along the imaginary 
$\omega$-axis in the usual way, letting $\omega=i\xi$, 
\begin{equation}
\beta_j(i\xi,k) \equiv i\kappa_j(\xi,k)=
i\sqrt{\varepsilon_j(i\xi)\xi^2/c^2+k^2},
\end{equation}
and noting that the expression in the brackets is real on 
the imaginary axis. We see that the regularized stress tensor is 
uniform across the layer. Although expected on invariance
grounds, this is not a trivial result and, as is clear from the 
derivation in the Appendix, it is due to cancellation of the 
$z$-dependent terms in the electric and magnetic contributions 
to $\tilde{T}_{j,zz}$ irrespective of the dielectric properties
of the surrounding stacks. 

Knowing the force, the Casimir energy ${\cal E}_j$ in the 
layer can be calculated using
\begin{equation}
f_{j-}=-f_{j+}=\frac{\partial {\cal E}_j}{\partial d_j},
\label{fEc}
\end{equation}
with the condition that ${\cal E}_j\rightarrow 0$ for
$d_j\rightarrow\infty$. From Eqs. (\ref{Trf})
and (\ref{f}), we find
\begin{eqnarray} 
{\cal E}_j&=&\hbar{\rm Im}\int_0^\infty\frac{d\omega}{2\pi}
\int\frac{d^2{\bf k}}{(2\pi)^2}\sum_{q=p,s}\ln{D_{qj}}(\omega,k)
\nonumber\\
&=&\frac{\hbar}{(2\pi)^2}\int_0^\infty d\xi
\int_0^\infty dkk\sum_{q=p,s}
\ln{D_{qj}}(i\xi,k).
\label{Ej}
\end{eqnarray} 
This equation, as well as that for the force [combined
Eqs. (\ref{f}) and (\ref{Trf})], agrees in form with the 
corresponding result of Zhou and Spruch \cite{zhou} derived 
using the (surface) mode summation method and starting from a 
simple model of a purely dispersive multilayer. However, 
in this work the contributions of all (propagating and evanescent) 
modes are naturally taken into account on an equal footing
through the Green function. Furthermore, since the Green 
function employed refers to a general absorbing multilayer, 
so do the obtained results except, of course, for the region 
where the Casimir force is calculated. 

The Casimir energy and force vary in a stepwise fashion across 
the multilayer and we end this section by pointing out a 
relationship that exists between their values in two different 
layers, say, layers ($j$) and ($l$). Indeed, assuming that $l>j$, 
for example, and using recursion relations for the reflection 
coefficients given by Eq. (\ref{rijk}), one may prove that the 
following relation exists between the $D$ functions for the 
layers \cite{tomas}:
\begin{equation}
D_{ql}(1-r^q_{j/l} r^q_{j-} e^{2i\beta_j d_j})=
D_{qj}(1-r^q_{l/j} r^q_{l+} e^{2i\beta_l d_l}).
\label{DjDl}
\end{equation}
Combining this with Eq. (\ref{Ej}), we find that the respective 
Casimir energies are related according to 
\begin{eqnarray} 
{\cal E}_l&=&{\cal E}_j+\frac{\hbar}{(2\pi)^2}\int_0^\infty 
d\xi\int_0^\infty dkk\times\nonumber\\
&&\sum_{q=p,s}\ln\left[\frac{1-r^q_{l/j}(\omega,k)r^q_{l+}(\omega,k)
e^{2i\beta_ld_l}}
{1-r^q_{j/l}(\omega,k)r^q_{j-}(\omega,k)
e^{2i\beta_jd_j}}\right]_{\omega=i\xi}.
\label{ElEj}
\end{eqnarray} 
A similar relation is obtained for the forces in two layers, but 
the resulting expression is not particularly illuminating unless 
$\varepsilon_j=\varepsilon_l$. Such a situation arises, for example, 
when a planar object is embedded in a planar cavity. In this case, 
we find
\begin{eqnarray} 
f_{l-}&=&f_{j-}+\frac{\hbar}{2\pi^2}\int_0^\infty d\xi
\int_0^\infty dkk\kappa\sum_{q=p,s}\frac{1}{D_{qj}(i\xi,k)}\times
\nonumber\\
&&\left[\frac{1-r^q_{j/l}(\omega,k)r^q_{j-}(\omega,k)
e^{2i\beta d_j}}{1-r^q_{l/j}(\omega,k)r^q_{l+}(\omega,k)
e^{2i\beta d_l}}-1\right]_{\omega=i\xi},
\label{fljD}
\end{eqnarray}
where $\beta$ ($\kappa$) is the perpendicular wave vector in both 
layers. 

\section{Discussion}
Most of the previously obtained results for the Casimir force 
and energy in a specific planar configuration are recovered from 
the results derived in the preceding section simply by specifying 
the corresponding reflection coefficients and material parameters. 
Thus, for example, the results for the three-layer 
($\varepsilon_1,\varepsilon_3,\varepsilon_2$) configuration considered 
by Lifshitz \cite{lifshitz} are obtained letting 
$\varepsilon_j=\varepsilon_3$, $r^q_{j-}\rightarrow r^q_{31}$ and
$r^q_{j+}\rightarrow r^q_{32}$, where $r^q_{ij}$ are single-interface 
reflection coefficients given by Eq. (\ref{sic}), and the results for 
the five-layer ($\varepsilon_4,\varepsilon_1,
\varepsilon_3,\varepsilon_2,\varepsilon_5$) configuration 
considered by Zhou and Spruch \cite{zhou} are obtained letting 
$\varepsilon_j=\varepsilon_3$, $r^q_{j-}\rightarrow r^q_{314}$ and 
$r^q_{j+}\rightarrow r^q_{325}$, where the three-layer reflection 
coefficients are obtained from recurrencies Eq. (\ref{rijk}).
Similarly, the results for the system consisting of two identical 
slabs, recently considered by Matloob and Falinejad \cite{matloob3}, 
are obtained letting $\varepsilon_j=1$ and $r^q_{j\pm}\rightarrow r^q$, 
where $r^q$ are the reflection coefficients of a symmetrically bounded 
slab [see Eq. (\ref{rt}) below], etc.
Specially, the Casimir force and energy in a (dispersive) planar 
cavity formed by two ideally reflecting (conducting) slabs are 
obtained with $r^q_{j-}r^q_{j+}=1$. 

We also note that these equations correctly reproduce the corresponding 
results which emerge from the 1D considerations. Indeed, taking only 
the $k=0$ contribution in Eq. (\ref{Trf}), we find from the
first line in that equation, for example,
\begin{equation} 
\tilde{T}^{1D}_{j,zz}=-\frac{2\hbar}{\pi}{\rm Re}
\int_0^\infty d\omega k_j(\omega)\frac{1-D_j(\omega)}
{D_j(\omega)},
\end{equation}
where $D_j(\omega)\equiv D_{qj}(\omega,0)$ [Eq. (\ref{Dj})] is the 
same for both polarizations. With a simple algebra, this 
equation can be rewritten as
\begin{eqnarray} 
\tilde{T}^{1D}_{j,zz}&=&\frac{\hbar}{\pi}
\int_0^\infty d\omega k_j(\omega)\times\nonumber\\
&&\left[1-
\frac{1-|r_{j-}(\omega)r_{j+}(\omega)|^2}
{|1-r_{j-}(\omega)r_{j+}(\omega)e^{2ik_j(\omega)d_j}|^2}\right],
\label{Trf1D}
\end{eqnarray}
which is in accordance with the Casimir force obtained 
by several authors for the respective systems they considered 
\cite{kupi1,kupi2,jaekel,matloob,esqu1,esqu2}.

Owing to the recursion relations which the generalized Fresnel 
coefficients satisfy, the obtained results can be applied to more 
complex systems with planar symmetry. As an application of the 
theory, we illustrate this by deriving  the Casimir force on a 
dielectric, or a metallic, slab 
[dielectric function $\varepsilon_s$, thickness $l$] in 
a cavity [dielectric function $\varepsilon$] with 
realistic mirrors [reflection coefficients $r^q_1$ 
and $r^q_2$], as depicted in Fig. 2. 
The force on the slab $f=f_{2-}-f_{1-}$ in this configuration can 
be calculated from Eq. (\ref{fljD}). The function $D_{q1}$ 
[Eq. (\ref{Dj})] is straightforwardly obtained letting $r^q_{1-}=r^q_1$ 
and $r^q_{2+}=r^q_2$ and using Eq. (\ref{rijk}) to determine 
the reflection coefficients $r^q_{1+}$. We find (the 
polarization index $q$ is omitted)
\begin{equation}
D_1=1-r_1\left(r+\frac{t^2r_2e^{2i\beta d_2}}
{1-rr_2e^{2i\beta d_2}}\right)e^{2i\beta d_1},
\end{equation}
where $r=r_{1/2}=r_{2/1}$ and $t=t_{1/2}=t_{2/1}$ are 
Fresnel coefficients for the slab. 
\begin{figure}[htb]
\begin{center}
\leavevmode
\hbox{%
\epsfxsize=8.6cm
\epsffile{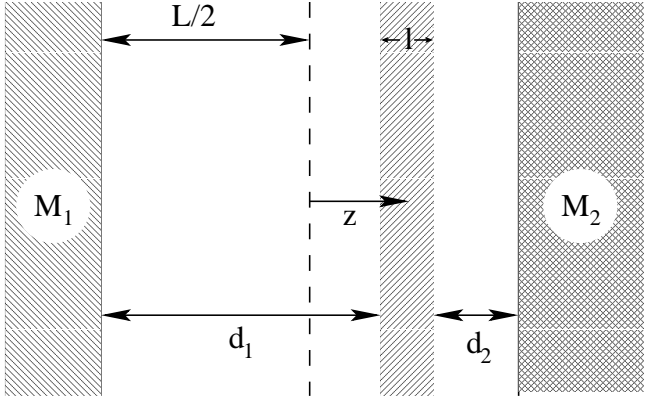}}
\end{center}
\caption{A dielectric slab in a planar cavity shown schematically. 
The arrow indicates the direction of the force on the slab.}
\end{figure}
\noindent
This gives
\begin{eqnarray} 
f&=&\frac{\hbar}{2\pi^2}\int_0^\infty d\xi\int dkk\kappa\times
\nonumber\\
&&\sum_{q=p,s}\left[\frac{r(r_2e^{2i\beta d_2}-
r_1e^{2i\beta d_1})}{N}\right]^q_{\omega=i\xi},\nonumber\\
&N&=1-r(r_1e^{2i\beta d_1}+r_2e^{2i\beta d_2})\nonumber\\
&&+d(r^2-t^2)r_1r_2e^{2i\beta (d_1+d_2)},
\label{f12}
\end{eqnarray}
where the expression in the brackets is to be calculated for 
$q$-polarization. Using Eqs. (\ref{rrel}) and (\ref{sic}), $r$ and $t$
can be further expressed entirely in terms of the reflection 
coefficient for the cavity-slab interface $\rho=(1-\eta)/(1+\eta)$
[where $\eta^p=\varepsilon\beta_s/\varepsilon_s\beta$ 
and $\eta^s=\beta_s/\beta$] as 
\begin{equation}
r=\rho\frac{1-e^{2i\beta_sl}}
{1-\rho^2e^{2i\beta_sl}},\;\;\;
t=\frac{(1-\rho^2)e^{i\beta_sl}}
{1-\rho^2e^{2i\beta_sl}}.
\label{rt}
\end{equation}
Note that for a perfectly conducting 
($\varepsilon_s\rightarrow \infty$) slab, we have $r^p=-r^s=1$ and
$t^q=0$.

The force $f$, as given by Eq. (\ref{f12}), may be positive or 
negative, depending on the dielectric properties of the slab 
and cavity mirrors as well as on the position of the slab. One may 
easily verify that this equation gives the correct result for the 
force on a perfectly conducting plate in an empty cavity with 
ideally reflecting walls. Indeed, since in this case $t^2=0$, 
N factorizes and Eq. (\ref{f12}) splits to ($r^2=rr_1=rr_2=r_1r_2=1$)
\begin{equation}
f=f_-(d_2)-f_-(d_1),
\label{fd12}
\end{equation}
where
\begin{eqnarray} 
f_-(d)&=&\frac{\hbar}{\pi^2}\int_0^\infty d\xi
\int_0^\infty dkk\frac{\kappa}{e^{2\kappa d}-1}
\nonumber\\
&=&\frac{\hbar}{3\pi^2c^3}\int_0^\infty d\xi\xi
\frac{\frac{d}{d\xi}\left[\sqrt{\varepsilon(i\xi)}\xi\right]^3}
{e^{2\sqrt{\varepsilon(i\xi)}\xi d/c}-1}
\label{f(d)}
\end{eqnarray}
is the force on the left mirror of a dispersive ideal cavity
[cf. Eqs. (\ref{f}) and (\ref{Trf}), with the index $j$ dropped].
The second line here is obtained upon a partial integration 
over $\xi$ and upon calculating the $\xi$-derivative of the integral 
over $k$ (see Ref. \cite{schaden}). For the empty cavity 
[$\varepsilon(i\xi)=1$], the integrals in Eq. (\ref{f(d)})
become elementary giving the well-known result 
\begin{equation}
f=\frac{\pi^2 \hbar c}{240}
\left(\frac{1}{d^4_2}-\frac{1}{d^4_1}\right),
\label{f012}
\end{equation}
according to which the plate is attracted to the closer cavity 
mirror. For a partially transmitting plate, the vacuum-field 
fluctuations in regions (1) and (2) of the cavity are no longer 
independent of each other and considerable deviations from 
the above simple result may occur especially for realistic 
cavity mirrors. Clearly, in this case, in order to explore the 
combined effect of the nearby walls on a planar (nano)object, 
one must analyze  Eq. (\ref{f12}) numerically. 
 
\section{Summary}
Using the properties of the macroscopic field operators appropriate 
for dissipative systems and a convenient Green function for a 
multilayer, in this work we have obtained general results for the 
Casimir force and energy applicable to local layered absorbing 
systems. We have also established a relationship between 
the Casimir force (and energy) in two different layers and, as an 
application of the theory, calculated the Casimir force on a 
dielectric slab in a realistic planar cavity.

\acknowledgments{This work was supported by the Ministry of Science 
and Technology of the Republic of Croatia under contract 
No. 00980101.}

\appendix
\section*{Green function}
Denoting the (conserved) wave vector parallel to the system 
surfaces by ${\bf k}=(k_x,k_y)$, we write the wave vector of 
an rightward (leftward) propagating wave in an $l$th layer
as ${\bf K}^\pm_l={\bf k}\pm\beta_l\hat{\bf z}$, where
\begin{equation} 
\beta_l=\sqrt{\tilde{k}^2_l-k^2}=\beta_l'+i\beta_l'',\;\;\;
\beta_l'\geq 0,\;\;\beta_l''\geq 0.
\end{equation}
With this notation, the Green function dyadic for the scattered field 
in the $j$th layer reads \cite{tomas}
\begin{eqnarray}
&&\tensor{\bf G}^{\rm sc}_j({\bf r},{\bf r}';\omega)=
\frac{i}{2\pi}\int\frac{d^2{\bf k}}
{\beta_j}e^{i{\bf k}\cdot({\bf r}_\parallel-{\bf r}_\parallel')}
\sum_{q=p,s}\frac{e^{i\beta_j d_j}}{D_{qj}}\xi_q\times\nonumber\\
&&\left\{r^q_{j-}e^{i\beta_j z_-}\hat{\bf e}^+_{qj}({\bf k})
\left[\hat{\bf e}^+_{qj}(-{\bf k})e^{-i\beta_jz'_+}+
r^q_{j+}\hat{\bf e}^-_{qj}(-{\bf k})e^{i\beta_jz'_+}\right]\right.
\nonumber\\&&\left.+r^q_{j+}e^{i\beta_j z_+}\hat{\bf e}^-_{qj}({\bf k})
\left[\hat{\bf e}^-_{qj}(-{\bf k})e^{-i\beta_jz'_-}+
r^q_{j-}\hat{\bf e}^+_{qj}(-{\bf k})e^{i\beta_jz'_-}\right]\right\},
\label{greensc}
\nonumber
\end{eqnarray}
\[D_{qj}=1-r^q_{j-} r^q_{j+} e^{2i\beta_j d_j},\;\;\;
\xi_p=1,\;\;\xi_s=-1,\]
\[z_-\equiv z,\;\;z_+\equiv d_j-z,\;\;0\leq z\leq d_j,\]
\begin{equation}
\hat{\bf e}^\mp_{pj}({\bf k})=\frac{1}{\tilde{k}_j}(\pm\beta_j
\hat{\bf k}+k\hat{\bf z}),\;\;\;\hat{\bf e}^\mp_{sj}({\bf k})=
\hat{\bf k}\times\hat{\bf z}\equiv\hat{\bf n},
\end{equation}
where $r^q_{j\pm}\equiv r^q_{j/n(0)}$
are, respectively, the transmission and reflection coefficient
of the upper (lower) stack of layers bounding the layer $(j)$.
Clearly, for the outmost layers, $l=n\;(0)$, we have
$r^q_{n+}=0$ and $r^q_{0-}=0$. Also, one must let 
$d_n\;(d_0)=0$ since these quantities appear only formally. 
The remaining Fresnel coefficients satisfy 
\begin{mathletters}
\label{rrel}
\begin{equation}
r^q_{i/j/k}=r^q_{i/j}+\frac{t^q_{i/j}t^q_{j/i}r^q_{j/k}
e^{2i\beta_jd_j}}{1-r^q_{j/i}r^q_{j/k}e^{2i\beta_jd_j}},
\label{rijk}
\end{equation}
\begin{equation}
t^q_{i/j/k}=\frac{t^q_{i/j}t^q_{j/k}e^{i\beta_jd_j}}
{1-r^q_{j/i}r^q_{j/k}e^{2i\beta_jd_j}}=
\frac{\beta_i}{\beta_k}t^q_{k/j/i},
\label{tijk}
\end{equation}
\end{mathletters}
and, for a single $i-j$ interface, reduce to
\begin{equation} 
r^q_{ij}=\frac{\beta_i-\gamma^q_{ij}\beta_j}
{\beta_i+\gamma^q_{ij}\beta_j},\;\;\; 
t^q_{ij}=\sqrt{\gamma^q_{ij}}(1+r^q_{ij}),
\label{sic}
\end{equation}
where $\gamma^p_{ij}=\varepsilon_i/\varepsilon_j$
and $\gamma^s_{ij}=1$, respectively. 

Performing the derivations indicated in Eq. \ (\ref{GB}) and using 
\begin{equation}
{\bf K}^\pm_j({\bf k})\times\hat{\bf e}^\pm_{qj}({\bf k})=
\tilde{k_j}\xi_q\hat{\bf e}^\pm_{q'j}({\bf k}),\;\;\;p'=s,\;\;s'=p,
\end{equation}
we find that $\tensor{\bf G}^{B,\rm sc}_j({\bf r},{\bf r}';\omega)$
is given by Eq. (A2) multiplied by $-\tilde{k}_j^2$ and with
$\hat{\bf e}^\pm_{qj}\rightarrow\hat{\bf e}^\pm_{q'j}$.
Noting that the equal-point Green function dyadics consist only of 
diagonal elements, we easily find
\begin{eqnarray} 
&&\tensor{\bf G}_j^{\rm sc}({\bf r},{\bf r};\omega)=
\frac{i}{2\pi\tilde{k}_j^2}\int\frac{d^2{\bf k}}{\beta_j}
\times\\
&&\left\{\hat{\bf k}\hat{\bf k}\frac{\beta_j^2}{D_{pj}}
\left[2r^p_{j-}r^p_{j+}e^{2i\beta_j d_j}-r^p_{j-}e^{2i\beta z_-}
-r^p_{j+}e^{2i\beta z_+}\right]\right.\nonumber\\ 
&&+\hat{\bf n}\hat{\bf n}\frac{\tilde{k}_j^2}{D_{sj}}
\left[2r^s_{j-}r^s_{j+}e^{2i\beta_j d_j}+r^s_{j-}e^{2i\beta_j z_-}
+r^s_{j+}e^{2i\beta_j z_+}\right]\nonumber\\
&&\left.+\hat{\bf z}\hat{\bf z}\frac{k^2}{D_{pj}} 
\left[2r^p_{j-}r^p_{j+}e^{2i\beta_j d_j}+r^p_{j-}e^{2i\beta_j z_-}
+r^p_{j+}e^{2i\beta_j z_+}\right]\right\},\nonumber
\end{eqnarray}
and $\tensor{\bf G}^{B,\rm sc}_j({\bf r},{\bf r};\omega)$
is given by this equation multiplied by $\tilde{k}_j^2$ and with
$p\leftrightarrow s$. The traces
$G^{\rm sc}_{j,\parallel}({\bf r},{\bf r};\omega)$ and
$G^{B,\rm sc}_{j,\parallel}({\bf r},{\bf r};\omega)$ 
can be easily recognized from these equations and one has, 
for example,
\begin{eqnarray}
&&\tilde{k}_j^2\left[G^{\rm sc}_{j,\parallel}({\bf r},{\bf r};\omega)-
G^{\rm sc}_{j,zz}({\bf r},{\bf r};\omega)\right]=
\frac{i}{2\pi}\int\frac{d^2{\bf k}}{\beta_j}
\times\\
&&\left\{\frac{\beta_j^2}{D_{pj}}
\left[2r^p_{j-}r^p_{j+}e^{2i\beta_j d_j}-r^p_{j-}e^{2i\beta z_-}
-r^p_{j+}e^{2i\beta z_+}\right]\right.\nonumber\\ 
&&+\frac{\tilde{k}_j^2}{D_{sj}}
\left[2r^s_{j-}r^s_{j+}e^{2i\beta_j d_j}+r^s_{j-}e^{2i\beta_j z_-}
+r^s_{j+}e^{2i\beta_j z_+}\right]\nonumber\\
&&\left.-\frac{k^2}{D_{pj}} 
\left[2r^p_{j-}r^p_{j+}e^{2i\beta_j d_j}+r^p_{j-}e^{2i\beta_j z_-}
+r^p_{j+}e^{2i\beta_j z_+}\right]\right\},\nonumber
\end{eqnarray}
while $G^{B,\rm sc}_{j,\parallel}({\bf r},{\bf r};\omega)-
G^{B,\rm sc}_{j,zz}({\bf r},{\bf r};\omega)$ is given by this 
equation with $p\leftrightarrow s$. Adding these two quantities, 
one finds that the curly bracket in Eq. \ (\ref{Tr}) is equal to
\begin{eqnarray}
\{\ldots\}&=&-8\pi i\int\frac{d^2{\bf k}}{(2\pi)^2}\beta_j\times\nonumber\\
&&\left[\frac{r^p_{j-}r^p_{j+}e^{2i\beta_j d_j}}{D_{pj}}
+\frac{r^s_{j-}r^s_{j+}e^{2i\beta_j d_j}}{D_{sj}}\right].
\label{cb}
\end{eqnarray}


\begin{references}

\bibitem{casimir}H. B. G. Casimir, Porc. K. Ned. Akad.
Wet. {\bf 51}, 793 (1948).

\bibitem{lifshitz0} E. M. Lifshitz, Zh. Eksp. Teor. Fiz.
{\bf 29}, 94 (1995) [Sov. Phys. JETP {\bf 2}, 73 (1956)].

\bibitem{schwinger}J. Schwinger, L. L. DeRaad, Jr., and K. A. Milton,
Ann. Phys. (N. Y.) {\bf 115}, 1 (1978).

\bibitem{milonni}P. W. Milonni, {\it The Quantum Vacuum.
An Introduction to Quantum Electrodynamics} (San Diego, 
Academic, 1994) Chap. 7.

\bibitem{bordag} For new developments in the Casimir effect 
and an extensive list of references see M. Bordag, U. Mohideen, 
G. L. Klimchitskaya,
and V. M. Mostepanenko, Phys. Rep. {\bf 353}, 1 (2002).


\bibitem{lamoreaux}S. K. Lamoreaux, Phys. Rev. Lett. {\bf 78},
5 (1997); {\it ibid} {\bf 81}, 5475(E) (1998).

\bibitem{mohideen}U. Mohideen nad A. Roy, Phys. Rev. Lett. {\bf 81},
4549 (1998).

\bibitem{bressi}G. Bressi, G. Carugno, R. Onofrio, and G. Ruoso,
Phys. Rev. Lett. {\bf 88}, 041804 (2002)

\bibitem{lamoreaux1}S. K. Lamoreaux, Phys. Rev. A {\bf 59},
R3149 (1999).

\bibitem{lambrecht}A. Lambrecht, and S. Reynaud, Phys. Rev. Lett.
{\bf 84}, 5672 (2000).

\bibitem{kampen}N. G. van Kampen, B. R. A. Nijboer, and K. Schram,
Phys. Lett. {\bf 26A}, 307, (1968).

\bibitem{schram}K. Schram, Phys. Lett. {\bf 43A}, 283, (1973).

\bibitem{zhou}F. Zhou and L. Spruch, Phys. Rev. A {\bf 52},
297 (1995).

\bibitem{klim}G. L. Klimchitskaya, U. Mohideen, and 
V. M. Mostepanenko, Phys. Rev. A {\bf 61}, 062107 (2000).

\bibitem{brown}L. S. Brown and G. J. Maclay, 
Phys. Rev. {\bf 184}, 1272 (1969).

\bibitem{ll}L. D. Landau and E. M. Lifshitz, {\it Electrodynamics
of Continuous Media}, (Pergamon, Oxford, 1991). Chap. 9.

\bibitem{brevik}See, e.g., I. Brevik and G. Einvoll, 
Phys. Rev. D {\bf 37}, 2977 (1988). 

\bibitem{kupi1} D. Kupiszewska, Phys. Rev. A {\bf 46},
2286 (1992).

\bibitem{kupi2}D. Kupiszewska nad J. Mostowski, 
Phys. Rev. A {\bf 41}, 4636 (1990).

\bibitem{jaekel} M. T. Jaekel and S. Reynaud,
J. Phys. I (France), {\bf 1}, 1395 (1991); A. Lambrecht,
M. T. Jaekel, and S. Reynaud, Phys. Lett. A {\bf 225}, 188 (1997).

\bibitem{matloob}R. Mathloob, A. Keshavarz, and D. Sedighi,
Phys. Rev. A {\bf 60}, 3410 (1999).

\bibitem{esqu1}R. Esquivel-Sirvent, C. Villarreal, and
G. H. Cocoletzi, Phys. Rev. A {\bf 64}, 052108 (2001).

\bibitem{esqu2}R. Esquivel-Sirvent, C. Villarreal, W. L. Moch\'{a}n,
and G. H. Cocoletzi, phys. stat. sol. (b), {\bf 230}, 409 (2002).

\bibitem{matloob2} R. Mathloob,
Phys. Rev. {\bf 60}, 3421 (1999).

\bibitem{matloob3}R. Mathloob and H. Falinejad,
Phys. Rev. A {\bf 64}, 042102 (2001).

\bibitem{mochan}W. L. Moch\'{a}n, C. Villarreal, and 
R. Esquivel-Sirvent, e-Print arxiv: quant-ph/0206119.

\bibitem{welsch}T. Gruner and D.-G. Welsch, Phys. Rev. A {\bf 53}, 
1818 (1996); H. T. Dung, L. Kn\"{o}ll, and D.-G. Welsch, {\it ibid},
{\bf 57}, 3931 (1998); S. Scheel, L. Kn\"{o}ll, and D.-G. Welsch, 
{\it ibid}, {\bf 58}, 700 (1998).

\bibitem{matloob1}R. Matloob, R. Loudon, S. Barnett,
and J. Jeffers, Phys. Rev. A {\bf 52}, 4823 (1995);
R. Matloob and R. Loudon, {\it ibid} {\bf 53}, 4567 (1996);
R. Matloob, {\it ibid} {\bf 60}, 50 (1999). 

\bibitem{tomas}M. S.Toma\v s, 
Phys. Rev. A {\bf 51}, 2545 (1995).

\bibitem{jackson}J. D. Jackson, {\it Classical Electrodynamics},
2nd ed. (Willey, New York, 1975) Chap. 6.

\bibitem{lifshitz} E. M. Lifshitz and L. P. Pitaevskii,
{\it Statistical Physics}, Part 2, (Pergamon Press, Oxford, 1991) Ch 8.

\bibitem{schaden}M. Schaden, L. Spruch and F. Zhou, Phys. Rev. A {\bf 57},
1108 (1998).

\end{references}
\end{document}